\documentclass[prb,floatfix,showpacs,superscriptaddress,longbibliography,twocolumn]{revtex4-2}
\usepackage[USenglish]{babel}
\usepackage{amsmath,amssymb,amsfonts}
\usepackage{epstopdf}
\usepackage{epsfig} 
\usepackage{graphicx}
\usepackage[T1]{fontenc}
\usepackage{color}
\usepackage{url}
\usepackage{changes}

\usepackage[colorlinks=true,linkcolor=blue,citecolor=red]{hyperref}

\def\a{\alpha}              
\def\e{\varepsilon}          
       \def\l{\lambda}       
                    \def\s{\sigma}
        \def\o{\omega}   
       \def\D{\Delta}    \def\L{\Lambda}  
                     
        \def\O{\Omega}

\def\eg{\mbox{\it e.g.\ }}  
\def\=={\equiv}

\def\qed{\raise1pt\hbox{\vrule height5pt width5pt depth0pt}}

\def\cG0{{\cal G}_0} 
\def\cG{{\cal G}}

\def\up{\uparrow} \def\down{\downarrow} 
  
\def\bk{{\bf k}}
\def\bq{{\bf q}}
\def\bp{{\bf p}}

 \def\q{{\bf q}}

\def\cc{c^{\dagger}}
\def\ca{c^{\phantom{\dagger}}}

\def\bc{b^{\dagger}}
\def\ba{b^{\phantom{\dagger}}}

\def\D0{D^{(0)}}

\newcommand{\quave}[1]{\langle{#1}\rangle}

\begin{document}

\author{Giacomo Mazza}
  \email{giacomo.mazza@unipi.it}
 \affiliation{Dipartimento di Fisica, Universit\`a di Pisa, Largo Bruno Pontecorvo 3, 56127, Pisa, Italy}
\affiliation{Department of Quantum Matter Physics, University of
  Geneva, Quai Ernest-Ansermet 24, 1211 Geneva, Switzerland}

\author{Adriano Amaricci}
\affiliation{CNR-IOM, Istituto Officina dei Materiali,
Consiglio Nazionale delle Ricerche, Via Bonomea 265, 34136 Trieste, Italy}

\title{Strongly correlated exciton-polarons in twisted homo-bilayer heterostructures}

\begin{abstract}
We consider dressing of excitonic properties by strongly correlated electrons 
in gate controlled twisted homo-bilayer heterostructures.  The
combined effect of the moir\'e potential and the Coulomb interaction
supports the formation of different strongly correlated phases depending 
on the filling, including charge-ordered
metals or incompressible insulators at integer occupation. The
coupling between excitons and electrons results in a splitting of the
excitonic resonance into an attractive and a repulsive polaron
peak. Analyzing the properties of the exciton-polarons across the 
different phases of the system, we reveal
a discontinuous evolution of the spectrum with the formation of a double-peak 
structure in the repulsive polaron branch.
The double-peak structure emerges for non-integer fillings 
and it is controlled by the energy separation
between the quasi-particle states close to the Fermi level
and the high-energy doublons excitations.
Our results demonstrate that exciton-polarons carry a clear hallmark 
of the electronic correlations
and, thus, provide a direct signature of the formation of correlation
driven insulators in gate controlled heterostructures.
\end{abstract}

\maketitle

The concept of strong correlation is generally used to  
express the competition between the single- and the many-particles
character of the excitation spectrum,
which is often at the origin of emerging
collective phenomena in quantum many-body systems.
In strongly correlated materials this competition, quantified by the 
relative weight of the electronic kinetic energy and the interaction 
energy issued from the Coulomb repulsion, can drive a characteristic transition
from a metal to a Mott insulator~\cite{Imada1998RMP}.
Yet, more than the metal-insulator transition, a distinctive hallmark
of the strongly correlated nature of a material is the existence of
concurrent excitations at different energy scales. 
For instance, long-lived quasi-particles control the low-energy
properties of the system whereas incoherent many-body excitations, associated with doublons 
formation, dominate the higher end of the energy spectrum.
This reflects in experimental outcomes, for example 
by means of direct coupling of the correlated electrons with 
optical probes~\cite{basov_electrodynamics_correlated_materials_RMP,
comanac_optical,sutter_calcium_ruthenate,
  tamai_strontium_ruthenate}.

In recent years, heterostructures of twisted two-dimensional 
materials emerged as a new class of electronic systems with tunable 
correlations thanks to the combined effects of moir\'e
potentials and electrostatic doping~\cite{correleted_ins_tblg,sc_tblg}.
These systems offered a platform to observe a number 
of phenomena, including metal-insulator transitions, 
quantum criticality or unconventional superconductivity~\cite{correleted_ins_tblg,sc_tblg,Wang2020_correlated_electronic_phases,
Ghiotto2021_quantum_criticality,
tang_simulation_of_hubbard2020,
Chen2019Evidence_gate_tunable}.
At the same time, they open new possibilities for detecting signatures of 
strong correlation, which go beyond the direct spectroscopic response 
of the correlated electrons~\cite{xie_spectroscopic_signature_MATBG,yang_spectroscopy_trilayer}.

Unlike most correlated materials, heterostructures made of 
transition-metal dichalcogenide (TMD)s 
display a semiconducting character in which strongly bounded excitons 
may act as sensors of correlated electronic states in the heterostructures~\cite{shimazaki_nature2020,shimazaki_prx2021,
wang_RMP_excitons,mott_moire_excitons_2207,berkelbach_reichman_annual_reviews_2018,Yang_sensing_proximity}.
The coupling between excitons and the electrons 
injected by gating can give rise to exciton-polarons (EP), issuing from the dressing 
of the original exciton resonance by the electronic cloud. 
In this process new bound states form, known as trions or
  charged excitons~\cite{jason_neutral_charged_excitons2013,kin_fai_mak_trions2013,
efimkin_prb2017,efimkin_prb2021,
sidler_fermi_polarons_2017}.
In wideband semiconductors, where Fermi liquid description of conduction
electrons applies, the EPs display a blueshift proportional to the 
density of conduction electrons~\cite{efimkin_prb2017,efimkin_prb2021,
sidler_fermi_polarons_2017}. 
This description has been used to map the charging 
diagrams of the TMDs systems and hint at the formation
of strongly correlated phases, such as the Wigner crystal in a monolayer 
MoSe$_2$~\cite{Smolenski2021} or Mott incompressible states in 
twisted MoSe$_2$/hBN/MoSe$_2$ heterostructures~\cite{shimazaki_nature2020,shimazaki_prx2021}.
Nonetheless, information on the charge density alone may not be conclusive 
for determining the correlated nature of carriers. Moreover, the standard EP 
description becomes questionable when Fermi-liquid theory breaks down.

In this paper, we derive a theoretical description of the properties 
of strongly correlated EPs in a paradigmatic setup of twisted homo-bilayer 
heterostructures. We obtain a charging diagram characterized by 
Mott and Wigner-Mott incompressible phases, with a layer-by-layer  
filling behavior in agreement with the experiments in Ref.~[\onlinecite{shimazaki_nature2020}].
We show that the EP spectrum discontinuously 
evolves as a function of the electronic density and splits into low- 
and high-energy branches of excitons dressed, respectively, 
by the coherent quasi-particles and the incoherent doublons.
We predict that for non-integer fillings, the EP spectrum shows multiple 
resonances at distinct energy scales, highlighting a direct signature of the 
formation of a correlation-driven Mott insulator.

The rest of this work is organized as follows: in
Sec.~\ref{secModelMethods} we introduce a model describing electrons and
excitons dynamics in a twisted homo-bilayer heterostructure. In
Sec.~\ref{secChargingDiagram} we solve the interacting electrons
problem and discuss the formation of different correlated phases at
increasing filling. Based on this result, in
Sec.~\ref{secExcitonPolarons} we investigate the coupling of electrons
with excitons and show how the resulting hybrid exciton-polaron
states display  hallmarks of strong correlation effects in the their
spectral properties. Finally, in Sec.~\ref{secConclusions} we draw
some conclusions and possible perspectives  deriving from our
results.

\begin{figure}
\includegraphics[width=\linewidth]{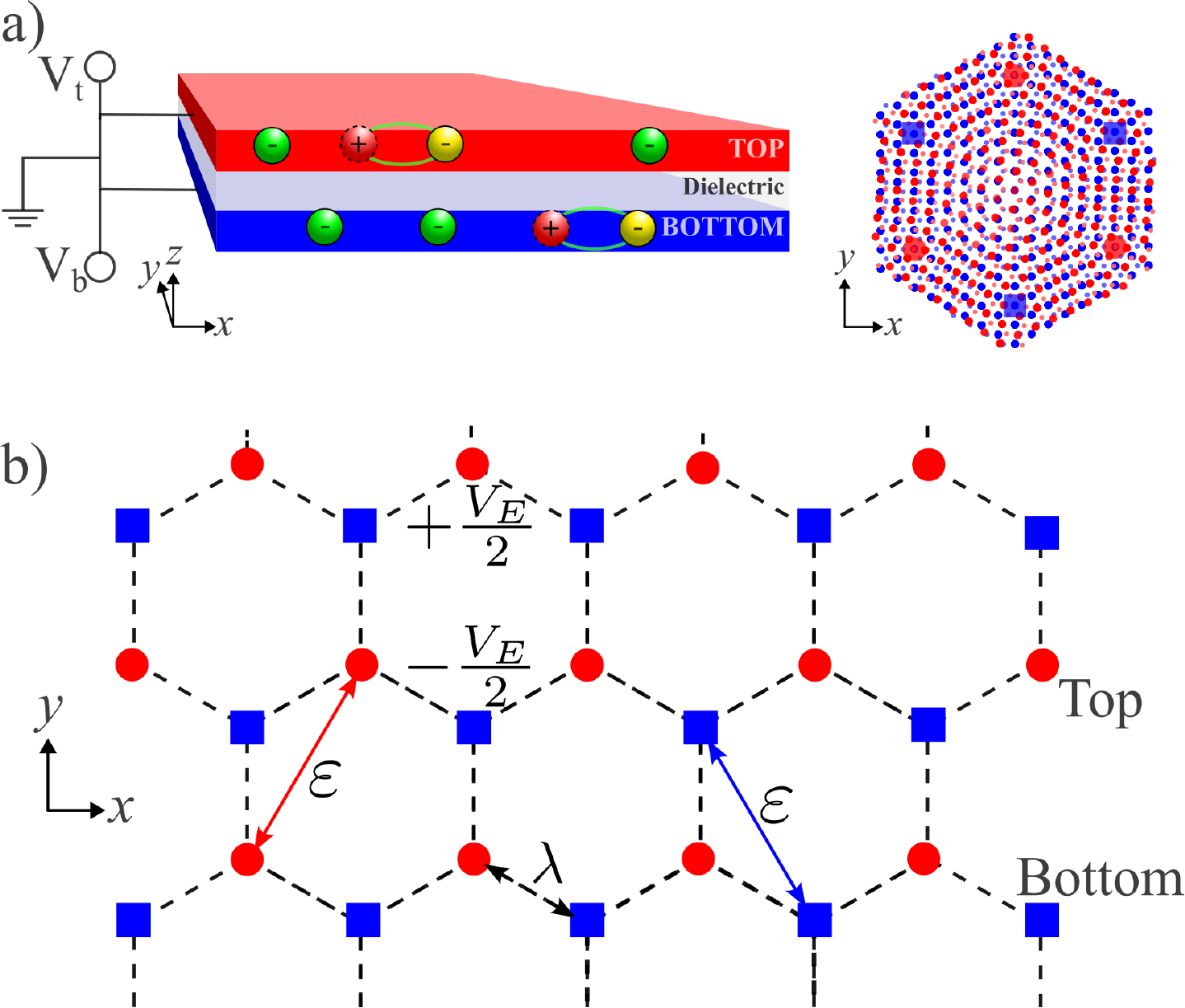}
\caption{ a) Right: Schematic representation of the homo-bilayer
  heterostructure. 
  Top (red) and Bottom (blue) layers are separated by a dielectric 
  spacer (gray).  Bound pairs of negative and positive 
  charges indicate intra-layer excitons.
  Green charges represents conduction electrons 
  controlled by top $V_t$ and bottom $V_b$ gates.
  Left: Twisted triangular lattices. Dots and squares
  highlight, respectively, the hexagonal lattice formed 
  by the MX and XM stacking.
  b) Top view of the lattice structure of the moir\'e Hamiltonian. 
  	Dots/squares sublattices correspond to orbitals localized 
  	on top/bottom layers. 
  	Arrows indicates the hopping matrix elements. 
  	The sublattice potential indicate the perpendicular bias $V_E$.}
\label{fig:fig0}
\end{figure}

\section{Model and Methods}\label{secModelMethods}
We consider two identical monolayers separated by a dielectric spacer (see Fig.~\ref{fig:fig0}).
The electron doping on each monolayer is controlled via top ($V_t$)
and bottom  ($V_b$) gates. The monolayers feature well-defined 
intra-layer exciton resonances.  A small twist angle gives rise to a moir\'e potential, 
with a periodicity much larger than the 
monolayer lattice parameter, leading to the 
formation of flat-bands in the mini-Brillouin zone, 
which due to the dielectric spacer are characterized
by a layer degree of freedom~\cite{shimazaki_nature2020,shimazaki_prx2021,
Zhang_homobilayer_prb2020,kuhlenkamp_tunable_topological_pseudospin_2209}.
Moreover, the moir\'e potential is assumed to be negligible for the intra-layer
exciton~\cite{Shabani2021NP,Linderalv2022}.
Its effect on the excitonic response reduces to 
the coupling between the intra-layer excitons and the electrons doped
into the moir\'e mini-bands.

We employ a phenomenological description of the 
electron-exciton problem, valid in the limit of 
low electronic density, corresponding to the electronic 
occupation of the first moir\'e mini-band with respect to available
 particle-hole excitations in the wide valence/conduction bands.
Our approach builds on the theory 
of composite many-body excitations  in quantum gases mixtures~\cite{Massignan_2014,
combescot_polarized_fermi_gases_PRL,chevy_original} which has been  successfully applied to the description of
trion absortion features in two-dimensional semiconductors~\cite{sidler_fermi_polarons_2017,efimkin_prb2021}. 
We describe intra-layer excitons as bosonic particles 
with dispersion $\o_{\bq}$ and we model their coupling  
with interacting electrons in terms of the following Hamiltonian: 
\begin{equation}
H = H_X + H_\mathrm{moir\acute{e}} + H_{e-X}.
\label{eq:H_elX}
\end{equation}
where $H_X = \sum_{\bq \a=t,b} \o_{\bq} \bc_{\bq \a} \ba_{\bq \a}$ represents the energy of the free exciton 
gases in the top and bottom layers, respectively $\a=t,b$.
$H_\mathrm{moir\acute{e}}$ describes the dynamics of the 
electrons in the moir\'e lattice and 
$H_{e-X}$ is the electron-exciton interaction.

Large scale electronic structure calculations for twisted homo-bilayers
TMDs~\cite{Zhang_homobilayer_prb2020} show that the low-energy moir\'e 
bands originate from orbital hybridization in a honeycomb 
lattice formed by the points in which the metal (M) atom of one 
layer is on top of a chalcogen (X) 
atom of the other, with opposite layer character for the MX or XM 
stacking (Fig.~\ref{fig:fig0}).
Based on that, we model the interacting electrons
as an extended Hubbard model on the honeycomb lattice, 
with the layer index corresponding to the sub-lattice one:
\begin{equation}
\begin{split}
H_\mathrm{moir\acute{e}} = & - \e \sum_{ij \in \mathrm{n.n.n.} } \sum_{\a \s} \cc_{i \s \a} \ca_{j \s \a} 
+ \lambda \sum_{ij \in \mathrm{n.n.} \s} \cc_{i \s t} \ca_{j \s b} + h.c. \\
& -\mu \sum_{i \s \a} n_{i \s \a}  
+ \frac{V_E}{2} \sum_{i \s} \left( n_{i \s t} - n_{i \s b} \right)
+ H_{int}.
\end{split}
\label{eq:H_el}
\end{equation}
where $\e$ and $\lambda$ are, respectively, the intra- and inter-layer
hopping between nearest neighbor sites,  $\mu=(V_t+V_b)/2$ is the chemical
potential controlling global filling and $V_E=V_{t}-V_{b}$ a
perpendicular bias appearing as an effective sublattice potential.
In Eq.~\ref{eq:H_el}, $\mathrm{n.n.}$ and $\mathrm{n.n.n.}$ symbols 
indicate, respectively, nearest and next-nearest neighboring sites. 
The interaction term contains both the local (intra-layer)
density-density interaction $U$ and 
the  nearest neighbor (inter-layer) one $U'$. 
\begin{equation}
H_{int} = U \sum_{i \a} n_{i \up \a} n_{i \down \a} 
+ U' \sum_{ij \in\mathrm{n.n.}} \sum_{\s \s'} n_{i  \s t} n_{j \s' b}.
\end{equation}
The local repulsion $U$ favors Mott localization whereas 
the non-local term $U'$ is responsible for Wigner 
crystallization, which in the present case corresponds
to the layer polarization.
Here, we focus on a regime in which both mechanisms 
are at play. We fix $U = 18\e$ and $U'= 0.25U$, 
and $\lambda=0.2\e$, with $\e=0.5\mathrm{meV}$ 
corresponding to flat-bands of width $W=4.5\mathrm{meV}$.
In the rest of this paper, we work at $T=0$ and we 
explicitly discard magnetic ordering.

Finally, we assume an effective electron-exciton contact 
attractive interaction deriving from a charge-dipole polarization 
mechanism~\cite{efimkin_prb2017,efimkin_prb2021,
sidler_fermi_polarons_2017}
\begin{equation}
  H_{e-X} = V_{e-X} \sum_{\bq \a=t,b}  
 \rho^{e}_{\bq \a} \rho_{-\bq \a}^X 
\end{equation}
where  $\rho_{\bq \a}^{e}  \equiv \sum_{\bk \s} \cc_{\bk-\bq \s \a} \ca_{\bk \s \a}$
and $\rho_{-\bq \a}^X \equiv \sum_{\bk} \bc_{\bk+\bq \a} \ba_{\bk \a}$
are, respectively,  the  electron and the exciton densities on the
layer $\a$.
In the following we set $V_{e-X}=-50~\mbox{meV}$.
This value is in line with the estimated short-range 
electron-exciton interaction in different TMDs~\cite{Fey2020PRB}, 
and it is found to correctly reproduce the order of magnitude of 
the experimentally observed polaron shifts~\cite{shimazaki_nature2020}.

\begin{figure}
\includegraphics[width=\linewidth]{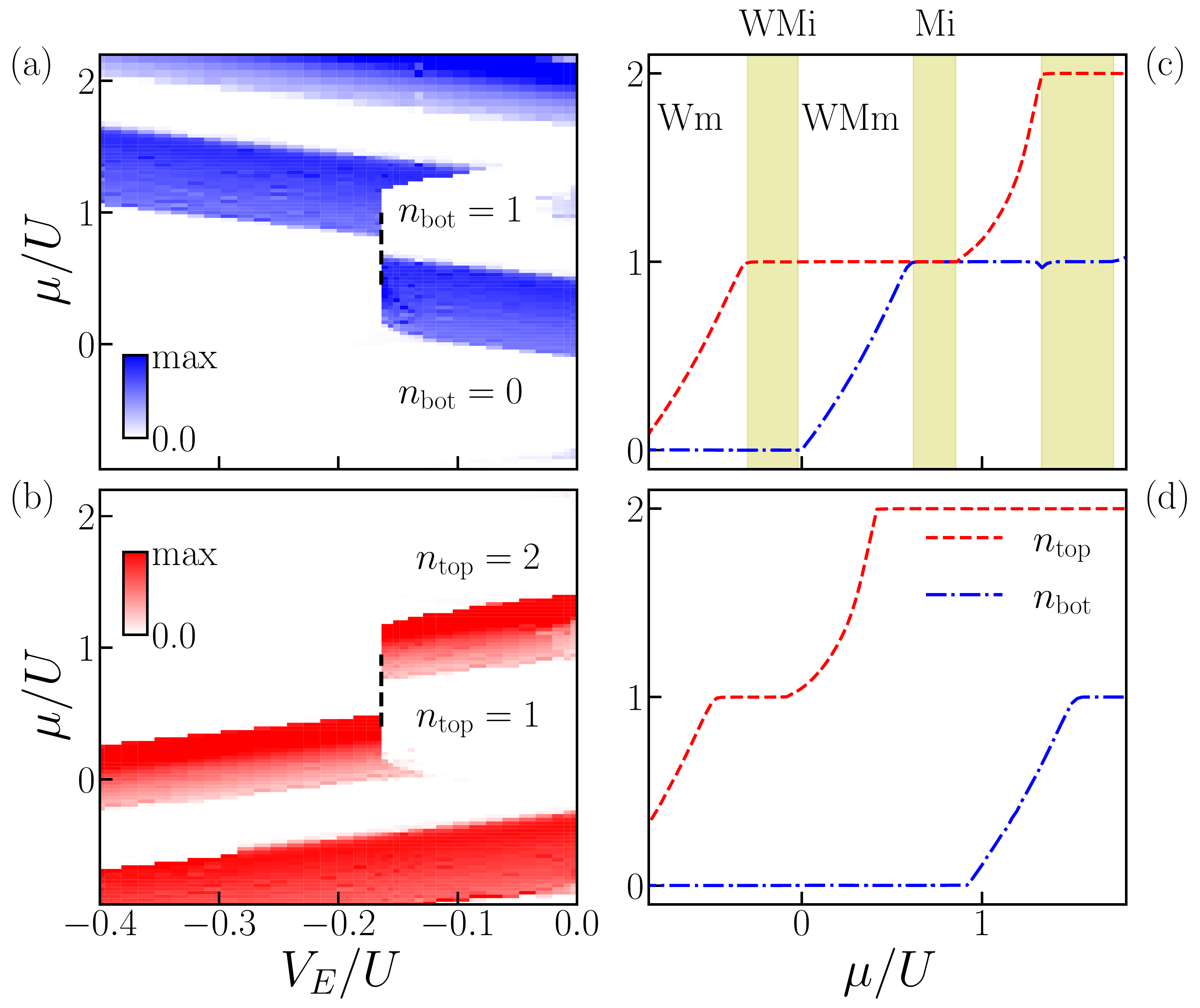}
\caption{
  Left Panels. Maps of the charge susceptibility $\chi_{\a}$,
  for the bottom (a) and top (b) layers, as a function of the chemical 
  potential $\mu$ and interlayer potential $V_E.$. 
  White areas correspond to charge incompressible regions realized for
  different fillings. The dashed lines mark the discontinuous transition between
  incompressible regions at different integer filling.    
  Right Panels. 
  Layer charge densities as a function of $\mu$ and fixed values of
  $V_E/U = -0.08$ (c) and $V_E/U=-0.25$ (d).
}
\label{fig:fig1}
\end{figure}

\section{Charging diagram}\label{secChargingDiagram}
We first discuss the charging diagram of the heterostructure.
We study the electronic Hamiltonian, Eq.~\ref{eq:H_el}, by 
using the Dynamical Mean Field Theory (DMFT)~\cite{dmft_review} with the 
Exact Diagonalization algorithm~\cite{EDIpack_2022} and decoupling 
of the non-local interaction~\cite{camjayi_natphys2008,amaricci_extended_prb2010,
merino_wigner_mott_2013,kapcia_prb2017,rademaker_2022}.
In Fig.~\ref{fig:fig1}, panels (a)-(b), we show maps of the layer charge 
susceptibilities  $\chi_{\a} \equiv \partial n_{\a} / \partial \mu $,
$n_{\a} = \sum_{i \s} \quave{\cc_{i \s \a} \ca_{i  \s \a}}$, as a
function of the chemical potential  $\mu$  and the interlayer bias $V_E$.
The layer susceptibilities display a characteristic checkerboard
pattern, alternating  layer-compressible regions ($\chi_{\a} \neq 0$)
to   layer-incompressible $\chi_{\a} = 0$ ones at  integer fillings, $n_\a = 0,\,1,\,2$.
The checkerboard pattern highlights the breaking of the layer symmetry
for $V_E \to 0$  due to Wigner crystallization.
The layer symmetry is restored in a region of the charging diagram 
characterized by charge densities simultaneously  plateauing at
half-filling $n_\mathrm{top}=n_\mathrm{bot} = 1$.
Panels (c)-(d) show cuts of the layer charge 
densities $n_\a$ as a function of $\mu$ for two distinct values of $V_E$.
The total filling $n_{\mathrm{tot}} = n_{\mathrm{top}} +
n_{\mathrm{bot}}$  increases through a series of compressible and 
incompressible configurations, indicating the metallic and the
insulating character of the heterostructure.
The plateaus at $n_{\a} = 0$ and $n_{\a} = 2$ correspond,
respectively, to completely empty and full bands while 
thise at half-filling  $n_{\a} = 1$ indicate the Mott localized phase. 

Depending on the charge configuration $\nu = (n_{\mathrm{top}},n_{\mathrm{bot}})$, 
we identify different types of correlated phases controlled by the gates.
The symmetric  configuration, $\nu=(1,1)$, corresponds to 
the homogeneous Mott insulator (Mi) in which electrons in both layers
simultaneously localize.
$\nu = (1,0)$ represents a Wigner-Mott insulator (WMi)
in which saturated charge ordering is concomitant 
to Mott localization of the charged layer~\cite{camjayi_natphys2008,amaricci_extended_prb2010,
merino_wigner_mott_2013}.
For any non-integer filling $x$, the $\nu = (1,x)$ configuration
denotes a Wigner-Mott metal (WMm) in which 
electrons in the half-filled layer localize  
whereas electrons in the partially filled layer remain 
itinerant~\cite{amaricci_extended_prb2010,merino_wigner_mott_2013,kapcia_prb2017}.
Finally $\nu= (x,0)$ corresponds to a Wigner metal (Wm) with saturated 
charge ordering and itinerant electrons in the partially filled layer.
At larger dopings, charge ordering and Mott localization 
coexists with completely filled bands, \eg $\nu = (2,1)$.
Increasing the bias, the Mi phase disappears in favor 
of a fully polarized band insulator, $\nu = (2,0)$.
Due to the first-order nature of polarization driven 
Mott transitions~\cite{werner_high_spin2007,amaricci_first_order,giacomo_field_driven},
the bias induced transition is first order, as indicated by 
dashed lines in panels (a)-(b).
Despite the simplicity of the model, 
the charging diagram highlights a characteristic layer-by-layer 
filling behavior which reproduces the salient features
the charging diagrams reported in [\onlinecite{shimazaki_nature2020},\onlinecite{shimazaki_prx2021}]
for twisted MoSe$_2$/hBN/MoSe$_2$ heterostructures.

\begin{figure}
\includegraphics[width=\linewidth]{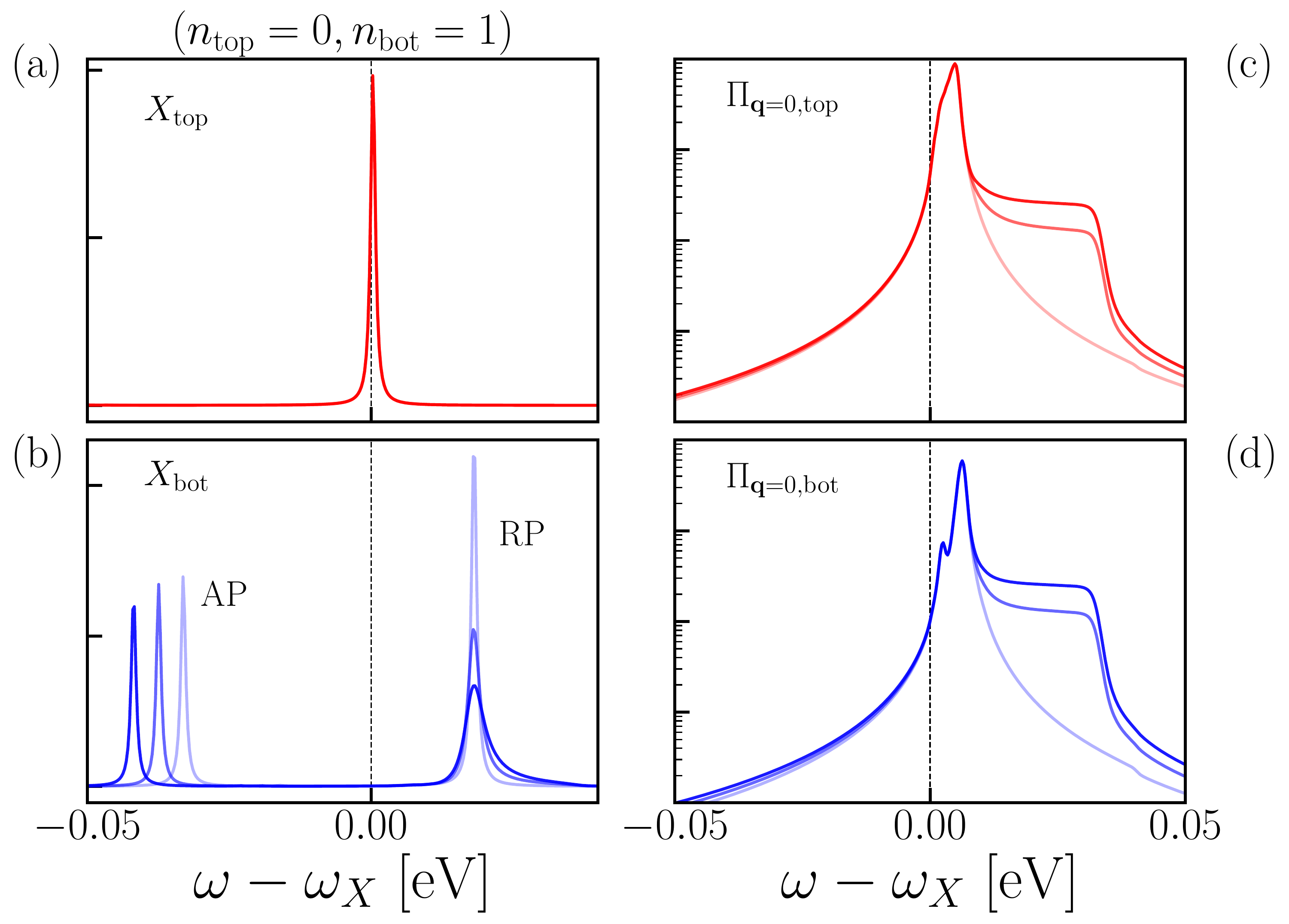}
\caption{Left panels. 
Spectral function for top (a) and bottom (b) 
interlayer excitons corresponding to the charging 
configuration $\nu=(0,1)$. Frequency is measured with 
respect to the energy of the bare exciton.
Right panels. Imaginary part of the electron-exciton pairing 
susceptibility at $\bq=0$ for top (c)
and bottom (d) layers. 
In each panel, the different curves correspond 
to increasing phenomenological background 
of electronic spectral density added in the calculations.
From light (moir\'e electrons only) 
to dark (moir\'e electrons with phenomenological background) curves.
In panel (a) the three curves are indistinguishable.
}
\label{fig:fig2}
\end{figure}

\section{Exciton-polarons}\label{secExcitonPolarons}
We now consider dressing of the excitonic properties
caused by coupling to the above strongly correlated electronic phases.
For low density the effect of the electrons onto the intra-layer
excitons can be treated perturbatively.
We are interested in the evolution of the $\bq=0$ excitonic spectral
function as probed by optical spectroscopy~\cite{Amelio2021PRB},
i.e.  $X_\a(\o) = -\tfrac{1}{\pi} \mathrm{Im}~D_{\bq=0
  \a}(\o) $, where $D_{\bq \a}(\o) =\left[\omega+ i\eta - \o_{\bq } -
\Sigma^X_{\bq \a}(\o)\right]^{-1}$ 
is the excitonic Green's function expressed in terms of the
excitonic self-energy $\Sigma^X_{\bq \a}$ function. 
The latter  can be evaluated from  the vertex function in 
the particle-particle channel $ \L_{\bp \s\a}$ as~
\footnote{
  In the self-energy we neglect the Hartree shift contribution that
  would produce a featureless red-shift of the exciton level. Indeed,
  we assume this contribution is counterbalanced by other effects,
  e.g. increase of the exciton energy due to screening, which are
  beyond the scope of the phenomenological model.}
\begin{equation}
\Sigma^X_{\bq\a}(i \O_n) =   \sum_{\bp \s} 
T \sum_{i \o_n} G_{\bp-\bq,\s\a}(i \o_n) \L_{\bp \s\a}(i\o_n + i\O_n). 
\label{eq:self-energy}
\end{equation}
We compute the vertex $\L_{\bp\s\a}$ in the  ladder
approximation~\cite{Massignan_2014,combescot_polarized_fermi_gases_PRL,Randeira_bound_states_1989}. 
\begin{equation}
\L_{\bp \s\a}(i\o_n)  =  V_{e-X} \frac{V_{e-X} \Pi_{\bp \s\a}(i \o_n)}{1-V_{e-X} \Pi_{\bp \s\a}(i \o_n)}
\label{eq:vertex_function}
\end{equation}
and the kernel $\Pi_{\bq\s\a}$ is the exciton-electron pairing susceptibility
\begin{equation}
\Pi_{\bq \s\a}(i \o_n) = - \sum_{\bp'} T \sum_{i \o_l} G_{\bp-\bp' \s\a}(i\o_l) D_{\bp' \a}(i\o_n-i\o_l).
\label{eq:pairing_pi}
\end{equation}
In these expressions $i \O_n$ and $i \o_{n}$ refer, respectively,
to the bosonic and fermionic Matsubara frequencies, while 
$G_{\bk\s\a}$ is the interacting electronic Green's function.
We assume the electrons act as a bath for the 
excitons, namely their properties are not affected by 
photogenerated excitons. The electron propagator 
$G_{\bk\s\a}$ is computed in the DMFT approximation using a 
momentum independent self-energy~\cite{dmft_review}.
Moreover, we compute the pairing susceptibility, Eq.~\ref{eq:pairing_pi}, 
using the bare exciton propagator which we assume having a parabolic 
dispersion $\o_{\q} = \o_X + \frac{\hbar^2 \bq^2}{2 M_X}$.
In the following we set a cut-off for the exciton and
the electron momenta within the first moir\'e mini-band.

In Fig.~\ref{fig:fig2}, we show the excitonic 
spectral functions $X_{\a=t,b}$ of the two layers  for the 
$\nu=(0,1)$ configuration.
For the undoped layer, the spectral function reduces to the bare one
with a slightly blue shifted resonance due to the small hybridization 
$\l$ between the layers.
At finite doping, the excitonic resonance  
splits in two distinct peaks located at energies, 
respectively, lower and higher than the bare exciton frequency.
Such splitting stems from the formation of a pole in the vertex function
outside the electron-exciton continuum.
The lower energy peak lies below the exciton-electron continuum and 
represents a bound state, identified as the attractive polaron (AP). 
The higher energy peak is a scattering state,
known as the repulsive polaron (RP).

Polaronic features are known to depend on the energy cut-off 
of the electron-exciton continuum~\cite{chevy_original,efimkin_prb2017,sidler_fermi_polarons_2017}.
  Here, we test the dependence of our results with respect to 
  the choice of the cut-off, by adding a small broad background 
  of constant spectral density in the exciton-electron continuum 
  (see Fig.~\ref{fig:fig2} (c) and (d)).
In panel (b) of Fig.~\ref{fig:fig2}, we show that the
position of the AP peak moves with the cut-off whereas 
the RP is, up to a broadening, essentially cut-off independent. The RP
thus represents a genuine feature of the interaction between excitons
and the correlated electrons in the first mini-band.
In the following, we will restrict our attention to the RP peak as a
mean to monitor the effects of strong electronic correlation.

\begin{figure}
\includegraphics[width=\linewidth]{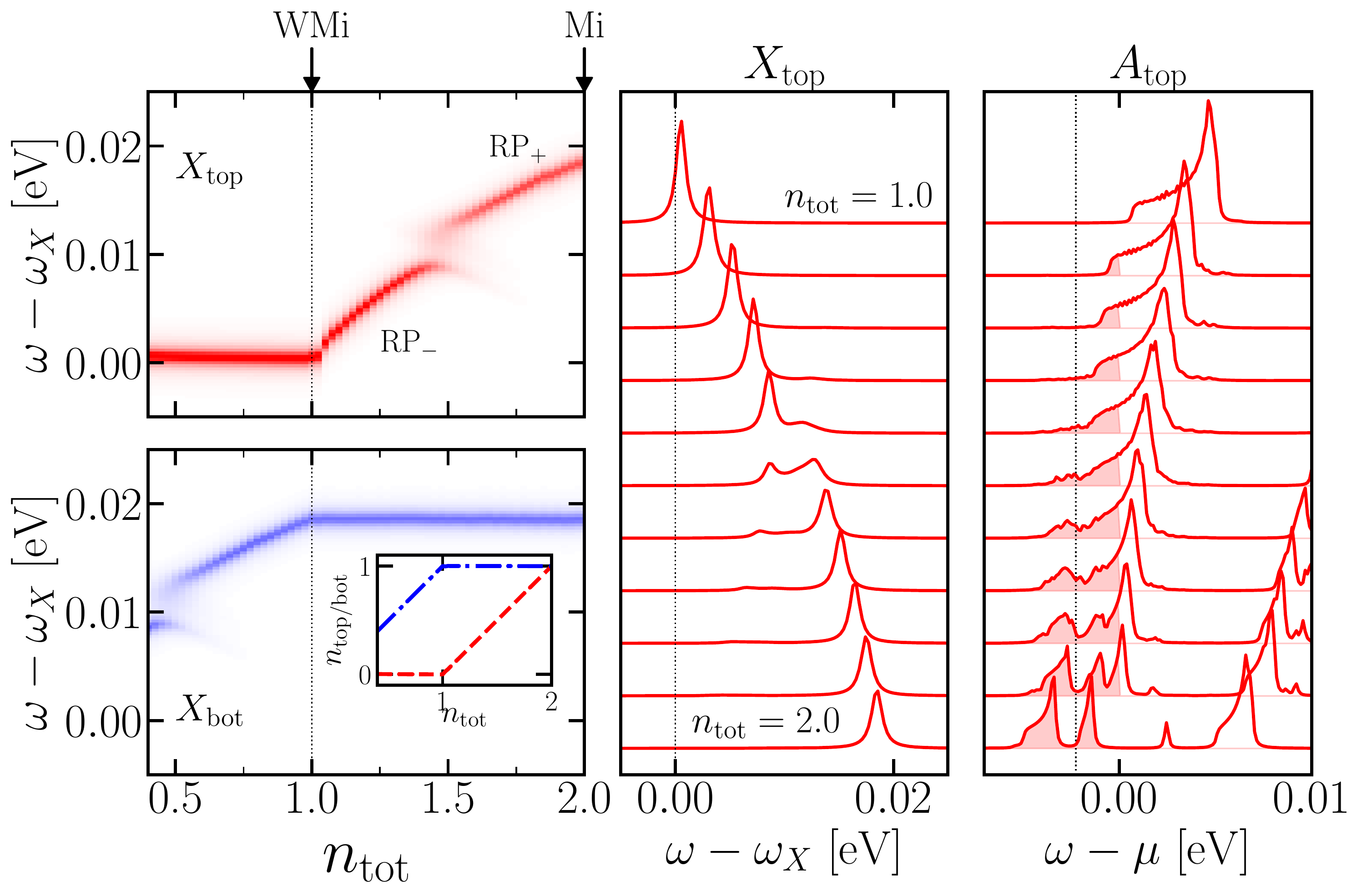}
\caption{Left panels. 
Color maps of the top and bottom excitons spectral functions 
as a function of the total filling $n_{\mathrm{tot}} $, the interlayer bias 
is $V_E/U=0.025$. 
Inset shows $n_{\mathrm{top}}$ (dashed line) 
and $n_{\mathrm{bot}}$ (dot-dashed line)
contributions to the total filling. 
The arrows and the vertical dotted line 
indicate the WMi and Mi phases at $n_{\mathrm{tot}} =1.0$ 
and $n_{\mathrm{tot}} =2.0$, respectively.
Center panel. 
Spectral density for the exciton in the top layer
at different values of the total filling from 
$n_{\mathrm{tot}} = 1.0$ to $n_{\mathrm{tot}} = 2.0$.
Vertical dotted line represents the bare exciton frequency.
Right panel. 
Electronic spectral density for the top layer 
for the same fillings in the center panel. 
Color filling highlights the occupied 
density of states. Vertical dotted line 
represents the energy scale $-(U-W)/2$ (see text) 
highlighting the formation of the lower Hubbard band.}
\label{fig:fig3}
\end{figure}
In Fig.~\ref{fig:fig3} we show the evolution of the RP resonance as a
function of the total filling through different correlated phases,
i.e. from the Wm ($n_\mathrm{tot} < 1$) to the Mi $(n_{\mathrm{tot}} = 2)$.
In all such cases, the RP resonance undergoes a blue-shift proportional to the 
density of the electrons, similar to what happens in 
the more conventional case of exciton-dressing by an electron-gas.

In the correlated electronic phases at non-integer fillings  
we observe an emerging structure in the RP peak.
We follow the top layer exciton in the 
WMm phase ($1< n_{tot} < 2$)  and observe that the 
RP resonance splits into two branches originating, respectively, 
from the empty (RP$_-$) and the half-filled layer
(RP$_+$). Remarkably, the two branches are not continuously connected
and appear as two separate peaks
  for the occupation value $n_{split} \approx 1.5$.

From the expression of the exciton self-energy, Eq.~\ref{eq:self-energy}, 
we relate the double-peak structure of the RP to the presence 
of distinct energy scales for the electronic excitations at 
non-integer fillings.
Indeed, frequency convolution in Eq.~\ref{eq:self-energy} 
implies that the excitonic scattering is sensible to the occupied 
electronic 
states, $\mbox{Im} \Sigma^{X}_{top}(\o) \sim - \theta(\o-\O_\L) A_{top}(\O_\L -\o)$
where $\O_\L$ is a characteristic resonance in the 
vertex function spectral density.
Due to the strong correlations, the electronic spectrum, $A_{top}$
 Fig.~\ref{fig:fig3}, feature a low-energy quasi-particle resonance close 
 to the Fermi level and higher-energy contributions 
 corresponding to the excitation of doublons. 

For small dopings, i.e. $n_{tot}\simeq 1$, the electron 
spectral weight is entirely composed of  low-energy
quasi-particle excitations, and the spectral density 
is equivalent to that of the bare triangular lattice.
As the filling increases, spectral weight starts to be transferred  
to the doublons, as highlighted in by the formation 
of the incoherent Hubbard band for energies 
$\o \lesssim -(U-W)/2 = -2.25~\mbox{meV}$, 
representing the energy edge of the Hubbard band. 
This reflects in two separated peaks in $\mbox{Im} \Sigma_{top}^X(\o)$.
The double-peak structure reveals two distinct 
scattering mechanisms, and manifests into two branches 
of solutions of the pole equation $\o-\o_X = \mbox{Re} \Sigma_{top}^X(\o)$.
We therefore link the RP$_-$ and RP$_+$ 
branches, respectively, with the quasiparticles and 
the  doublons excitations.  
The avoided crossing occurs when the relative 
weights of these two electronic excitations become comparable.
The splitting is found to increase proportionally to $U$.
We emphasize that the precise value of the 
non-integer filling $n_{split}$ smoothly depends on 
the model parameters, \eg the value of the 
electron-exciton interaction.

A similar trend can be observed for the bottom layer in the 
Wm phase $n_{tot} < 1.$  
We conclude that the RP$_-$ and RP$_+$ branches bear 
the hallmark of electronic correlations in the WMm and Wm phases.

\section{Conclusions}\label{secConclusions}
We derive a model to describe the coupling of interacting electrons
and exciton states in twisted homo-bilayer heterostructures.
In particular, by means of DMFT we solved the 
interacting electrons problem showing that the combination of Coulomb
repulsion and gate voltage leads to a characteristic charging diagram, 
featuring a checkerboard pattern of alternating correlated metallic
and insulating phases. 
Using a combination of numerical and analytic methods we treated the
electron-exciton coupling to analyze the properties of the EP hybrid
states, emerging from strongly correlated electrons dressing the
excitons in such systems.
We have shown that EP  resonances bear the hallmark of
strong correlation, namely the coexistence of low-energy
quasiparticles and high-energy incoherent excitations. We highlight
the discontinuous evolution of the exciton spectrum as filling is
varied between a Wigner metal and a Mott insulator. In particular, we
predict that, near semi-integer total filling, low- and high-energy
features equally participate in the formation of EP states, giving rise
to characteristic peaks separated by an energy 
that depends on the
interaction strength.
The direct consequence of our findings is that the genuine correlated nature
of either metallic or insulating phases in such systems can be
immediately identified by means of exciton spectroscopy. 
Our description can be extended to more specific materials modeling or
heterostructure setups for sensing of 
correlated electrons~\cite{Li_local_sensing_2021}. Investigating the
interplay of inter- and intra-layer interacting excitons~\cite{Amelio2022} is an
interesting research direction left for future work. 

\section{Acknowledgments}
We thank M.~Fabrizio for useful discussions. This work has been
supported by the Swiss National Science Foundation through an AMBIZIONE grant. 

\bibliography{biblio_exciton_polaron}

\end{document}